\documentclass[10pt,letterpaper]{article}
\usepackage{opex3}
\usepackage{color}

\begin{document}

\title{Simultaneous cooling of coupled mechanical oscillators using whispering gallery mode resonances} 

\author{Ying Lia Li, James Millen and P. F. Barker$^{*}$}

\address{Department of Physics \& Astronomy, University College London, WC1E 6BT, United Kingdom}

\email{$^*$p.barker@ucl.ac.uk} 



\begin{abstract}
We demonstrate simultaneous center-of-mass cooling of two coupled oscillators, consisting of a microsphere-cantilever and a tapered optical fiber. Excitation of a whispering gallery mode (WGM) of the microsphere, via the evanescent field of the taper, provides a transduction signal that continuously monitors the relative motion between these two microgram objects with a sensitivity of 3\,pm. The cavity enhanced optical dipole force is used to provide feedback damping on the motion of the micron-diameter taper, whereas a piezo stack is used to damp the motion of the much larger (up to $ 180\,\mu$m in diameter), heavier (up to $1.5\times 10^{-7}\,$kg) and stiffer microsphere-cantilever. In each feedback scheme multiple mechanical modes of each oscillator can be cooled, and mode temperatures below 10\,K are reached for the dominant mode, consistent with limits determined by the measurement noise of our system. This represents stabilization on the picometer level and is the first demonstration of using WGM resonances to cool the mechanical modes of both the WGM resonator and its coupling waveguide. 
\end{abstract}



\section{Introduction}

Optomechanical devices on the nano- and microscale are important tools, both for exploring fundamental quantum physics, and for technological applications. These systems, in which mechanical motion is controlled by light, can be used as exquisitely sensitive force \cite{geraci2010,kippenberg2015,geraci2015}, position \cite{rousseau06} or mass \cite{zhu12, hossein13} sensors. 

An important class of optomechanical systems exploit high optical quality factor whispering gallery mode (WGM) resonances ($\mathrm{Q}_{\mathrm{opt}} > 10^8$) in objects such as dielectric spheres or toroids. High-Q$_{\mathrm{opt}}$ WGM resonators \cite{haroche93, kimble98, rauschenbeutel09} have been used for microwave-to-optical conversion \cite{maleki03, yu09}, optical storage \cite{wang11}, frequency comb creation \cite{maleki08}, and single-molecule sensing in both air and liquids \cite{arnold08}. Strong optomechanical coupling can be achieved using the interaction between the optical WGM and {\em internal} mechanical modes \cite{kippenberg14}, which has been used to cool the radial breathing modes of a toroid to the ground state in a sideband-resolved regime \cite{riviere}. Unlike schemes which use optical forces such as radiation pressure to cool a cavity mirror or an object inside a cavity \cite{kleckner2006, corbitt2007}, WGM resonances offer the ability to cool and manipulate the center-of-mass (c.o.m.) motion of external objects placed in the evanescent field of the WGM cavity \cite{kippenberg2015, lin}. 

Many WGM experiments involve excitation using a tapered optical fiber. This form of coupling is prone to modulation due to the mechanical modes of the waveguide and coupling distance drifts. This extraneous noise is problematic for high-bandwidth sensor development \cite{krause} and is present in many types of optomechanical systems \cite{basarir}. Stabilization of the coupling junction has so far been limited to seismic bandwidths \cite{bowen12}.

The use of WGMs to cool and manipulate the {\em external} c.o.m. motion of the cavity that supports them has not been explored. An example of such a scheme has been proposed which utilizes narrow WGM resonances to cool the WGM cavity in a similar way to the Doppler cooling of atoms \cite{barkerwgm}. Cooling the c.o.m. of macroscopic objects is attractive for developing ultrasensitive closed-loop force sensors and exploration of quantum physics at large mass scales \cite{chen13,arndt14}. Recently, the mechanism for transducing the c.o.m. motion of a silica microsphere-cavity on the end of a silica pendulum (analogous to our system) has been observed and studied using light coupled into a WGM from a tapered optical fiber \cite{transduction}. This high resolution transduction at the taper-microsphere junction is dependent on coupling distance and detuning, and is seen as an important first step for implementation of active feedback cooling of either the microsphere resonator or the tapered fiber. Simultaneous damping of both these objects is an attractive feature for inertial sensing, as the taper stabilization minimizes drift, and damping of the microsphere-cantilever can extend the sensing range and decrease integration times \cite{gavartin12}.

In this article we use this transduction provided by a WGM of a microsphere-cantilever to feedback-cool the c.o.m. of the microsphere-cantilever and the tapered fiber to the noise floor. Light is coupled into the WGM via the tapered optical fiber, and its motion is feedback-cooled using the optical dipole force (effective mode mass $\approx$ 1.4\,$\mu$g). Simultaneously, a piezoelectric stack (PZT) attached to the heavier and stiffer microsphere-cantilever allows us to damp its motion as well (effective mode mass $\approx$ 20\,$\mu$g). We demonstrate cooling of many mechanical modes of both mechanical oscillators.

\section{Transduction of Thermal Motion}
\begin{figure}[h]
\centering
\includegraphics[width=12cm]{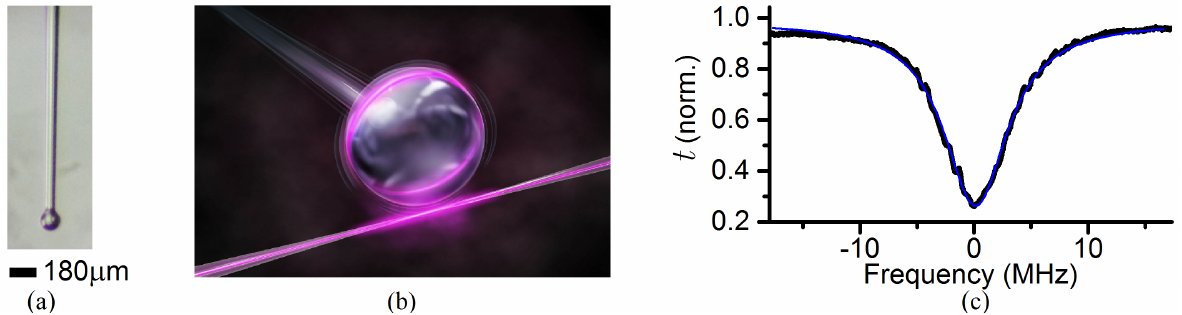}
\caption{\label{objects}(a) A microscope image of a typical microsphere-cantilever. (b) Artist's impression of the microsphere-cantilever, which is brought close to, but never touches, the tapered optical fiber underneath it. The evanescent fields of the WGM and tapered fiber are illustrated. (c) Monitoring the transmission $t$ through the taper as the laser frequency is scanned reveals a WGM resonance (black), fitted with a Lorentzian function (blue) with a FWHM of 6.8\,MHz.}
\end{figure}

To demonstrate c.o.m. cooling using WGMs we fabricate two oscillators from tapered optical fiber. The first oscillator is a dielectric microsphere on a cantilever, fabricated by melting the end of a fiber with a focused C$\mathrm O_{2}$ laser, which can produce microspheres $40\,\mu$m - $200\,\mu$m in diameter, see for example Fig.~\ref{objects}(a). A microsphere with a diameter of 177\,$\mathrm \mu$m is selected, which remains attached to the fiber stem 120\,$\mathrm \mu$m in diameter and 5.6\,mm long. The second oscillator consists of the tapered fiber which couples light into WGMs of the microsphere. The taper is produced in the standard way by pulling an optical fiber while it is heated, to create a 1\,$\mathrm \mu$m waist. In order to excite a WGM resonance of the microsphere, laser light is coupled into the tapered fiber which is brought within close proximity (less than 1\,$\mu$m) to the surface of the microsphere (Fig.~\ref{objects}(a), (b)). The laser frequency is scanned to locate a resonance (Fig.~\ref{objects}(c)). Typically, an upper limit of 80\% of the light propagating through the fiber is coupled into a WGM. The remaining 20\% loss is due to non-resonant scattering from the fiber. 

\begin{figure}[h]
\centering
\includegraphics[width=8.4cm]{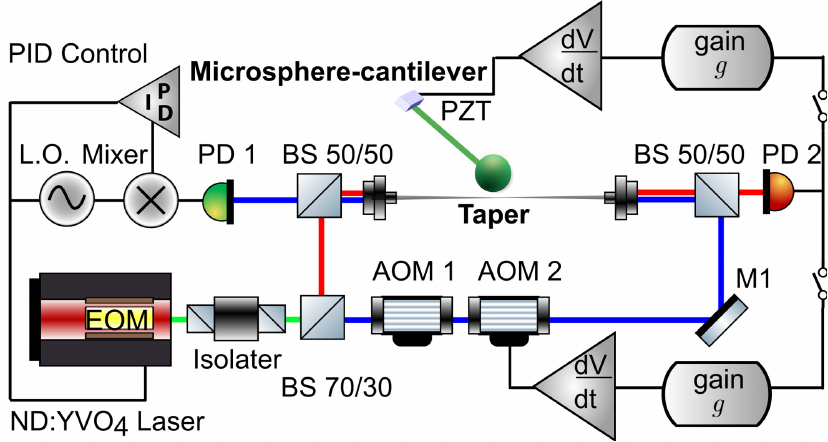}
\caption{\label{setup} A schematic of the transduction and feedback system. A 1064\,nm Nd:YV$\mathrm O_{4}$ laser with an intracavity electro-optic modulator (EOM) is tuned to excite the WGM. A PDH locking scheme is implemented, with sidebands created using an RF signal applied to the EOM. The light is split into a strong beam (blue), and a weak, red-detuned, transduction beam (red) that counterpropagate along the tapered optical fiber with fixed detuning. Feedback cooling is achieved by amplifying (increasing the gain, $g$) and differentiating the transduction signal from photodiode PD 2, which is sent to a piezo stack (PZT) supporting the microsphere-cantilever, and / or used to modulate the locking beam intensity using AOM 2. }
\end{figure}

\begin{figure}[h]
\centering
\includegraphics[width=12cm]{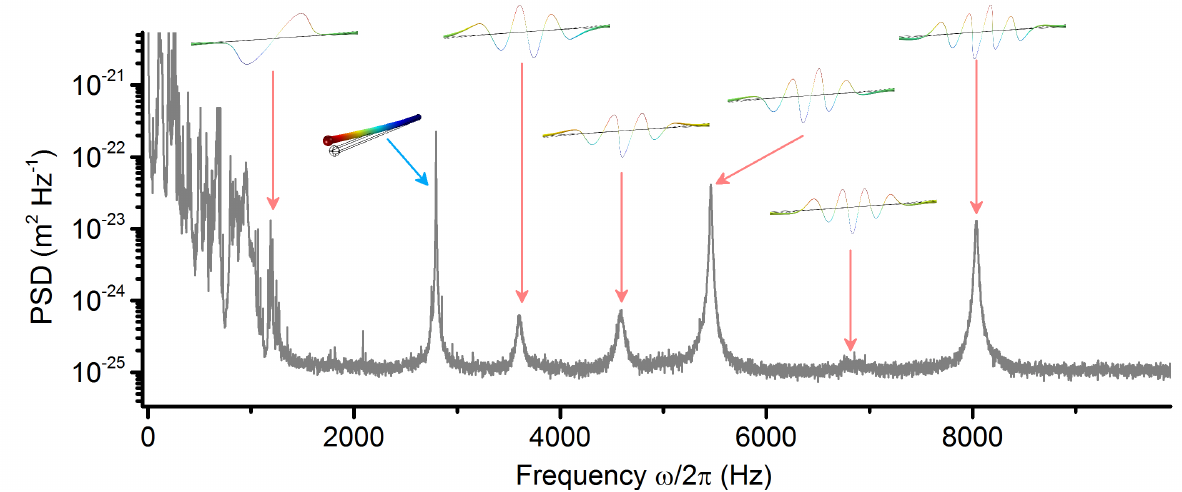}
\caption{\label{psdmodes}The full PSD from the transduction signal shows multiple mechanical modes belonging to the microsphere-cantilever or the tapered fibre which are identified by resonantly driving either oscillator. The predicted mode-shapes are shown as inset diagrams.}
\end{figure}

The optical layout is shown in Fig.~\ref{setup}. The tapered optical fiber is placed directly below the microsphere-cantilever and the system is within a vacuum chamber. 
We use a weak beam ($\approx 70\,\mu$W) propagating through the tapered fiber to measure the separation, $d$, between the taper and the microsphere, which we call the transduction beam.  The transduction results from changes in the coupling efficiency due to fluctuations of the coupling gap between the taper and microsphere, which is seen as a change in the transmission through the fiber. A strong beam ($\approx 300\,\mu$W) counterpropagates through the same fiber, and is used to control the position of the taper via the cavity enhanced optical dipole force (CEODF) \cite{painter} between it and the microsphere.  The strong beam is locked to the center of the WGM using Pound-Drever-Hall (PDH) locking or via thermal self locking \cite{carmon} when a large modulation of the light intensity to control the CEODF is required which destabilizes the PDH. The frequency of the transduction beam can be tuned with respect to the strong beam using acousto-optic modulators (AOM 1\&2).  The transmitted light is detected on a photodetector (PD 2) and the power spectral density (PSD) of the fluctuation of the transmitted light is used to extract the mechanical motion of both the microsphere-cantilever and the tapered fiber. Reducing the pressure from atmospheric to 1\,mbar enhances the mechanical quality factor $\mathrm{Q_{m}}$ of the taper by over a factor of 30, whereas $\mathrm{Q_{m}}$ for the much more massive microsphere-cantilever only increases by a factor of 1.2 over this range.
Using this setup, we find that the fundamental c.o.m. mode of the microsphere-cantilever is at approximately 2.8\,kHz, while a number of taper modes are observed between 0.3\,kHz - 15\,kHz, with prominent higher order peaks found between 3.8\,kHz - 8\,kHz due to their large mass participation factors. The modes are experimentally distinguished by resonantly driving either the taper or the microsphere-cantilever, with further evidence provided by modeling the system using the finite element modeling package COMSOL. The full PSD is presented in Fig.~\ref{psdmodes}.

\begin{figure}[h]
\centering
\includegraphics[width=12cm]{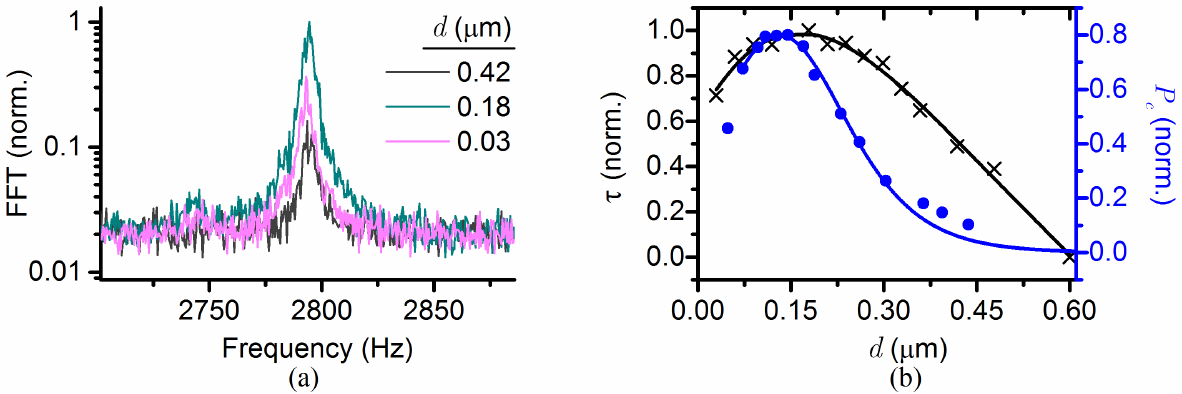}
\caption{\label{transduction1}(a) The FFT of the transduction signal, showing the motion of the microsphere-cantilever with $d = 0.42\,\mu$m, $0.18\,\mu$m, and $0.03\,\mu$m, illustrating the change in transduction, defined as the peak height above the background level. The FFT is normalized with respect to the maximum measured peak height. (b) The effect of the coupling distance, $d$, on the power coupled into the WGM, $P_{c}$, normalized to the input power propagating through the taper (blue), fitted with the on-resonance transmission relationship from \cite{painter}. Also shown is the transduction, $\tau$, of the fundamental mechanical mode of a microsphere-cantilever (black) when the transduction beam is red-detuned by 20 MHz using a WGM with 40\,MHz FWHM, fitted using reference \cite{transduction}.}
\end{figure}

Optimal feedback cooling requires maximizing the transduction signal $\tau$ above noise.  This signal varies with the taper-microsphere separation, $d$, and the detuning from the WGM resonance. We determine the optimal coupling distance by changing $d$ to maximize the amplitude of the detected mechanical motion above the noise floor, as shown in Fig.~\ref{transduction1}(a). The peak of this motion is defined as $\tau$, and the optimization of $\tau$ with $d$ is shown in Fig.~\ref{transduction1}(b). 
The optimum separation was found to be 150\,nm, which corresponds to the distance at which the maximum power is coupled into the WGM. These results are in very good agreement with recent work \cite{transduction}, where the authors show a maximum in transduction at the critical coupling distance. Previous studies revealed that red-detuning the transduction beam from resonance provided the strongest transduction \cite{transduction}, which we verified for this system by measuring the peak amplitude relative to the background level as the laser is tuned in frequency across the WGM \cite{lia}. This detuning sign-dependence indicates that both dispersive and dissipative changes in the coupling are important. Our transduction sensitivity can detect displacements as small as 3\,pm which was confirmed by calibration with the supporting PZT stack.

\section{Active Feedback Cooling}
We can cool both mechanical oscillators, individually or simultaneously. We perform active feedback cooling of the tapered fiber motion using the CEODF, first to demonstrate that optically cooling a micron sized object is achievable, and second, to show stabilization of the coupling junction through reducing the mechanical motion of the taper itself. This type of stabilization differs from that reported in reference \cite{bowen12} which prevents low frequency drifts ($< 20$\,Hz) in coupling distance. Modulation of the CEODF can cool the taper modes, but cannot efficiently actuate the heavier microsphere-cantilever. This is because the effective motional mass is over an order of magnitude lower for the taper modes ($\approx$ 1.4\,$\mu$g) compared to the microsphere-cantilever ($\approx$ 20\,$\mu$g). 
The transduction signal at the taper frequency is amplified and differentiated to provide a velocity dependent damping term that is used to modulate the power of the strong beam. The arising CEODF is attractive and given by \cite{painter}:
\begin{equation}
 F_{o}(d)=-\frac{8\psi\chi_{t}V_{t}\gamma_{e}(0)\mathrm{Q}_{\mathrm{opt}}^{2}P_{c}}{\Theta_{0}^{2}n_{t}^{2}V_{s}}\times\frac{e^{\psi d}}{\left(e^{\psi d}+\left(\gamma_{e}(0)\mathrm{Q}_{\mathrm{opt}}/\Theta_{0}\right)\right)^{3}} ,
\label{CEODF}
\end{equation}
where $V_{s}$ and $V_{t}$ are the effective mode volumes of the microsphere and taper respectively, $n=1.54$ is the refractive index of silica, $\chi_{t}=0.8$ is the effective susceptibility of the taper,  $\gamma_{e}(0)$ is the extrinsic (taper waveguide to WGM resonator) coupling rate at $d=0$, $\psi$ is the decay constant, $\Theta_{0}$ is the WGM resonance frequency in rad $\mathrm{s^{-1}}$ and $P_{c}$ is the power coupled into the WGM. We extrapolate $\gamma_{e}(0)=1.8\times10^{10}$\,rad\,s$^{-1}$ and $\psi=1.4\times10^{7}$\,m for a WGM resonance with an optical Q-factor of $6.1\times10^{5}$, following the procedure in \cite{painter}. The force between the taper and microsphere is predicted to be 28\,pN when using 100\,$\mu$W of light coupled into a WGM supported by a 100\,$\mu$m diameter microsphere, at the optimum coupling separation.

In the presence of feedback the PSD of the mechanical motion becomes:

\begin{equation}
S^{\mathrm{fb}}(\omega)=\frac{2\Gamma_{0}k_{B}T_{0}}{M_{\mathrm{eff}}}\frac{1}{(\omega_{0}^{2}-\omega^{2})^{2}+(1+g)^{2}\Gamma^{2}_{0}\omega^{2}},
\label{psd}
\end{equation}

where $\omega$ is the observed mechanical frequency, $\omega_{0}$ is the natural mechanical frequency, $\Gamma_{0}$ is the mechanical damping factor, $M_{\mathrm{eff}}$ is the effective mass of the resonator mode, $T_{0}$ is the undamped temperature, and $g$ is the feedback gain. The value of $g$ is proportional to the feedback signal which is differentiated in time to provide velocity damping. This feedback signal modulates the laser power which changes the power coupled to the WGM, $P_{c}$, in Eq.~\ref{CEODF}. When active feedback is applied, the mechanical motion is equivalent to that of an oscillator with a higher damping factor $\Gamma_{\mathrm{fb}}=(1+g)\Gamma_{0}$, and a lower temperature, $T_{\mathrm{fb}}$, where $T_{\mathrm{fb}}=(1+g)^{-1}T_{0}$.

\begin{figure}[h!]
\centering
\includegraphics[width=12cm]{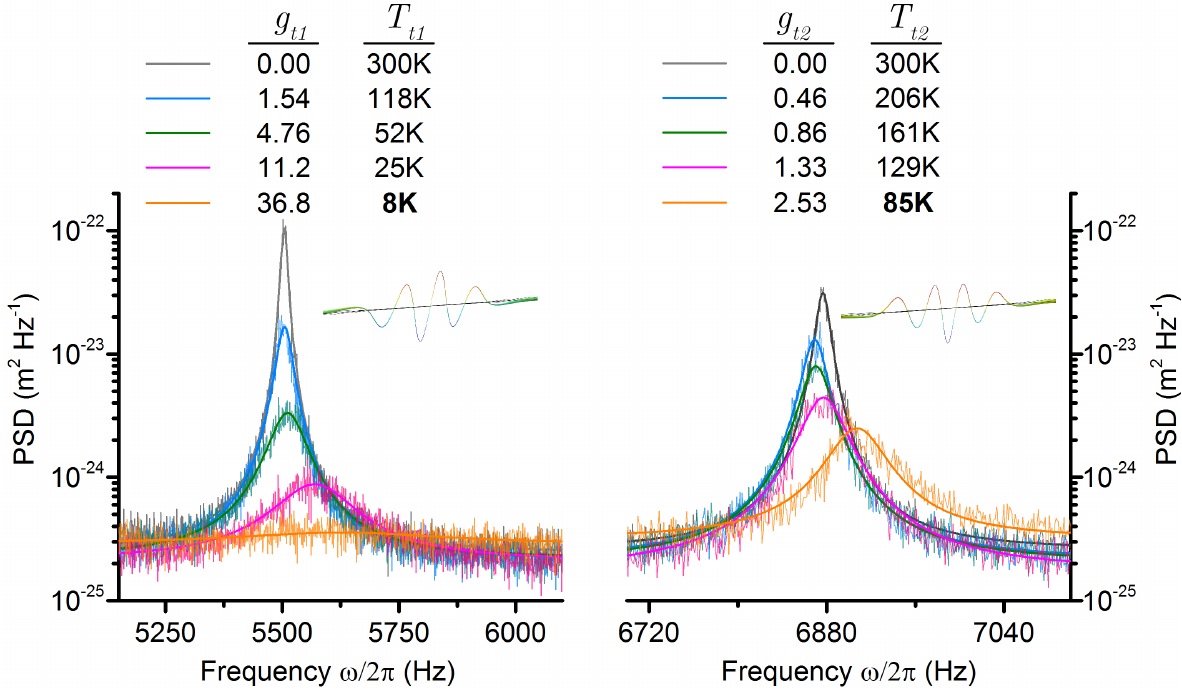}
\caption{\label{cooling} Simultaneous cavity enhanced optical dipole force cooling of two mechanical modes of the tapered optical fiber, obtained at a pressure of 0.5\,mbar. Representational mode shapes are shown. Each mode temperature is defined as $T_{t1}, T_{t2}$, at varying gain, $g_{t1}, g_{t2}$. Curves are fitted using Eq.~\ref{psd} to infer the mode temperatures, and the values of $T_{t1,t2}$ are found with less than 10\% uncertainty. The mechanical quality factor of each mode is $\mathrm{Q_{m1}}=380$ and $\mathrm{Q_{m2}}=510$. }
\end{figure}

In Fig.~\ref{cooling} is displayed a plot of the recorded PSD of two taper modes at 5.50\,kHz ($t1$) and 6.88\,kHz ($t2$), for different values of the feedback gain $g_{t1}, g_{t2}$. The feedback is onto the strong beam intensity, which modulates the CEODF. The modes are simultaneously cooled to 8\,K and 85\,K, respectively, at a pressure of 0.5\,mbar. Taper mode $t1$ has a different feedback phase relationship to $t2$, leading to different complex feedback gains $g_{t1}, g_{t2}$ for each mode, so both cannot be simultaneously cooled efficiently with the same signal. 

The second mechanical oscillator is the microsphere-cantilever. The large effective mass of this simple structure is a promising candidate for ultrasensitive acceleration detection, whose resolution can be quantified by its thermal noise equivalent acceleration $a_{th}=\sqrt{\frac{4k_{B}T\omega_{0}}{M_{\mathrm{eff}}\mathrm{Q_{m}}}}$. It is interesting to note that for these massive microsphere-cantilevers $\mathrm{Q_{m}}$ is only weakly effected by lowering the pressure, allowing ambient operation. The value of $a_{th}$ is unaffected by feedback cooling, as the ratio $\frac{T}{\mathrm{Q_{m}}}$ remains fixed. However, following the work of \cite{hosseini}, the use of periodic quiescence feedback cooling could offer an advantage for improving signal-to-noise ratios for the measurement of classical impulse forces.

\begin{figure}[h!]
\centering
\includegraphics[width=12cm]{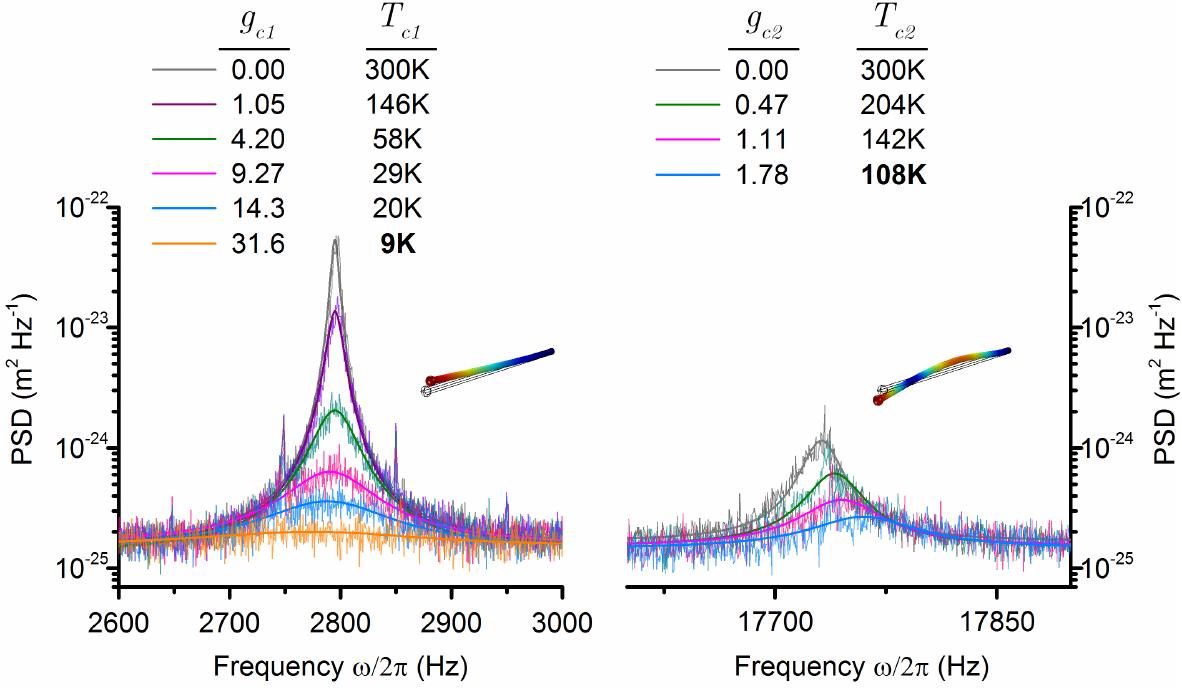}
\caption{\label{pztcooling}PZT feedback cooling of the fundamental mode of the microsphere-cantilever with $\mathrm{Q_{m1}}=370$, reaching $T_{c1}$ at varying feedback gains $g_{c1}$. The second mechanical eigenfrequency at 17.73\,kHz, with $\mathrm{Q_{m2}}=640$, is simultaneously cooled to $T_{c2}=108$\,K. Mechanical mode shapes are shown as inset diagrams. Data was taken at atmospheric pressure, with curves fitted using Eq.~\ref{psd} to infer the mode temperatures $T_{c1,c2}$, with uncertainties of less than 10\%.}
\end{figure}

To feedback cool the fundamental c.o.m. motion of the microsphere-cantilever we again use the transduction signal but apply the amplified and differentiated signal to a PZT supporting the clamped end of the microsphere-cantilever. The motion of the fundamental frequency at 2.80\,kHz ($c1$), as well as the second eigenfrequency at 17.73\,kHz ($c2$) can be detected, and in Fig.~\ref{pztcooling} we show the PSD for these modes as we apply feedback cooling. Cooled mode temperatures of $T_{c1}=9$\,K and $T_{c2}=108$\,K are reached for the fundamental mechanical frequency and the second eigenfrequency at the maximum gain, at atmospheric pressure. 

For both the optical cooling of the taper modes, and the PZT cooling of the microsphere-cantilever, we reach close to the cooling limit, as set by noise from electronics, photodetectors and laser noise. We note that the predicted lowest mode temperature is set by $T_{\mathrm{min}}=\sqrt{\frac{k \omega_{0} T_{0} S^{\mathrm{det}}}{k_{B}\mathrm{Q_{m}}}}$, where $k$ is the spring constant and $S^{\mathrm{det}}$ is the detection noise ($\approx 2\times10^{-26}$\,m$^{2}\mathrm{rad}^{-1}\mathrm{s}$). We place a lower bound of $T_{t1}\approx 8$\,K for the cooling of the taper mode at $\omega_{0}/2\pi=5.50$\,kHz ($k=1.7$\,Nm$^{-1}$, $\mathrm{Q_{m}}=380$), and $T_{c1}\approx 11$\,K for the microsphere-cantilever mode at $\omega_{0}/2\pi=2.80$\,kHz ($k=6.3$\,Nm$^{-1}$, $\mathrm{Q}_{m}=370$). The second mechanical eigenfrequency of the cantilever is cooled close to its respective limit of $T_{c2}\approx 80$\,K, representing effective broadband cooling for this particular oscillator. 

\begin{figure}[h!]
\centering
\includegraphics[width=8.4cm]{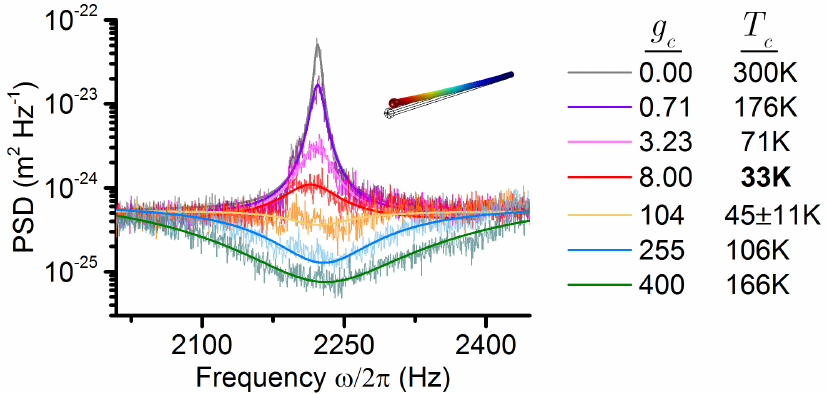}
\caption{\label{SQUASH} PZT cooling of a microsphere-cantilever with a lower $\mathrm{Q_{m}}=280$ than in Fig.~\ref{pztcooling} and higher noise floor, showing squashing with large feedback gain, $g_{c}$. The mechanical mode is shown as an inset. Data was taken at atmospheric pressure, with curves fitted using Eq.~\ref{psd} \& Eq.~\ref{psdmodified} to infer the mode temperatures, with less than 10\% uncertainty unless stated.}
\end{figure}

By using a different microsphere-cantilever, with similar dimensions, but a lower mechanical $\mathrm{Q_{m}}$ of 280 and a higher electronic noise floor, we show the effect of squashing \cite{hosseini, poggio} through high feedback gain in Fig.~\ref{SQUASH}(b). Under these conditions, increasing the gain beyond the cooling limit pushes the {\em measured} thermal noise spectra below the measurement noise. The feedback loop counteracts intensity fluctuations in the light field, which heats the actual cantilever motion as noise is imprinted on the oscillator. Calculation of the mode temperature therefore requires a modified PSD function \cite{hosseini, poggio}:

 \begin{equation}
S^{\mathrm{mod}}(\omega)=S^{\mathrm{fb}}(\omega)+S^{\mathrm{det}}\frac{(\omega_{0}^{2}-\omega^{2})^{2}+\Gamma^{2}_{0}\omega^{2}}{(\omega_{0}^{2}-\omega^{2})^{2}+(1+g)^{2}\Gamma^{2}_{0}\omega^{2}},
\label{psdmodified}
\end{equation}
where $S^{\mathrm{fb}}$ is defined in Eq.~\ref{psd}. The effective mode temperature for high gain was inferred using $T_{\mathrm{mod}}=\frac{T_{\mathrm{0}}}{1+g}+\frac{g^{2}}{1+g}\frac{k \omega_{\mathrm{0}} S^{\mathrm{det}}}{4k_{B}\mathrm{Q_{m}}}$, showing heating of the microsphere-cantilever motion at large values of $g$.

Ultimately, for microsphere-cantilevers of this size and mechanical $\mathrm{Q_{m}}$, improving the cooling limit can only be achieved by lowering the noise floor of our system, and optimizing the displacement resolution of the transduction by using narrower WGMs.

\begin{figure}[h!]
\centering
\includegraphics[width=12cm]{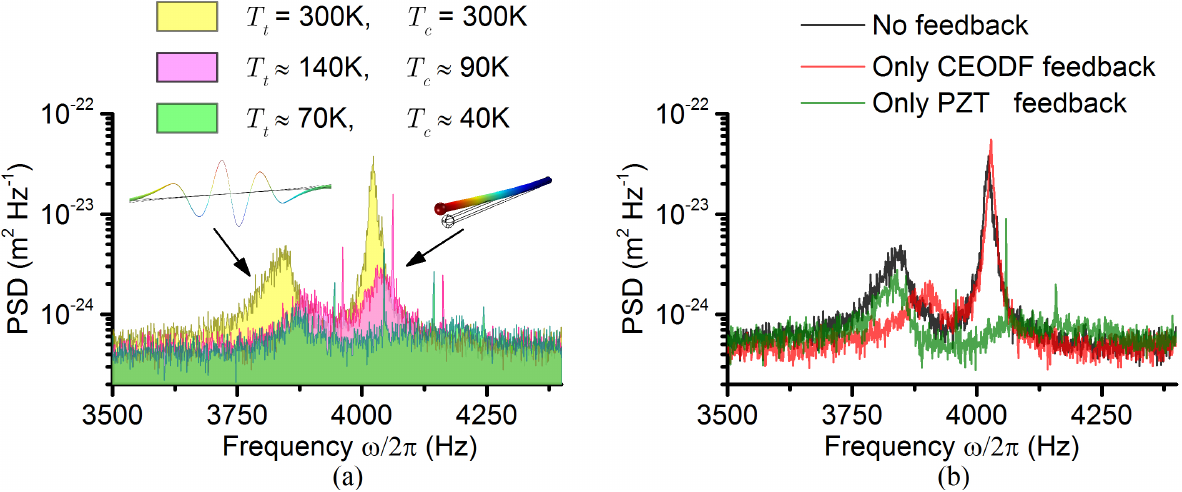}
\caption{\label{combined}(a) Simultaneous cooling of a taper mode at $T_{t}$ and the microsphere-cantilever mode at $T_{c}$ using both the PZT and the CEODF, performed at atmospheric pressure. Mode temperatures are approximated by integrating the area under the respective peaks. (b) Cooling with {\em only} the CEODF or the PZT scheme, each optimized for the taper mode and cantilever mode respectively, can influence the transduced PSD of the other oscillator.  }
\end{figure}

Finally, we show simultaneous operation of both feedback cooling schemes. The performance for simultaneous cooling of the taper mode at 5.5\,kHz (Fig.~\ref{cooling}) and the microsphere-cantilever mode at 2.8\,kHz (Fig.~\ref{pztcooling}) is not as efficient as the cooling of each mode separately. This occurs because the bandwidth of our feedback amplifiers is not narrow enough to isolate only one mechanical mode. This leads to crosstalk in the feedback signals sent to the PZT and for laser power modulation. Significant improvements in performance are possible by employing narrower filters that would completely isolate each mode. Instead, a different microsphere-cantilever and taper system is selected where the mechanical modes are less than 250\,Hz apart in frequency, and we remove any filters from the feedback scheme. We choose this system to demonstrate that the same transduction signal can be used to cool two separate mechanical oscillators when the elimination of cross-talk by using filters is not possible. The PSD of this system is presented in Fig.~\ref{combined}(a). 
Fitting with Eq. \ref{psd} can not be reliably applied at high gain, as the mechanical damping factor and the effective mode temperature no longer obey $(g+1)=\frac{\Gamma_{\mathrm{fb}}}{\Gamma_{\mathrm{0}}}=\frac{T_{\mathrm{0}}}{T_{\mathrm{fb}}}$. This is due to the interplay between the feedback schemes which rely on the transduction signal that measures {\em relative} displacement changes between these two seperate oscillators. Instead, we infer the cooled mode temperatures by integrating the area, and show cooling of both oscillators to less than 70\,K. Such a coupled system requires further investigation, but using experimental data shown in Fig. \ref{combined}(b) it is observed that each individual feedback scheme has an effect on the apparent mechanical damping factor and center frequency of the other oscillator. 

\section{Conclusion}
In conclusion, we have shown c.o.m. feedback cooling of two micron-scale, microgram, coupled mechanical oscillators using a WGM resonance. Mechanical modes belonging to the microsphere-cantilever and the tapered optical fiber used to excite this WGM can be individually cooled to below 10\,K, representing position stabilization on the picometer scale. The mechanical modes of the fiber can be troublesome for any hybrid WGM system using tapered fiber coupling \cite{krause}, and to our knowledge, direct cooling of these modes has not previously been demonstrated. Both oscillators can be cooled simultaneously, stabilizing the coupling distance, which enhances the potential to use the cooled system for ultraprecise closed-loop force sensing. Currently, our cooling scheme is limited by detection noise and the low mechanical $\mathrm{Q_{m}}$ of the oscillators. However, the ability to cool the c.o.m. motion of objects within this intermediate size and mass scale has not been fully investigated, and is attractive for creating macroscopic quantum objects. By stiffening the tapered fiber, or coupling light using a prism, the cavity enhanced optical dipole force could be used to cool the c.o.m. motion of a microsphere 10's $\mu$m in size, which could then be unclamped, i.e. levitated. Other cooling schemes for levitated particles, such as cavity or feedback cooling, would not work for such large objects. We note that when the laser input is locked to the WGM the transduction is purely dissipative and could be used for dissipative cooling of the microsphere taper system. Such dissipative cooling can in principle lead to ground state cooling even in the unresolved sideband regime \cite{clerk, pernice2009}.

\end{document}